\documentclass[conference]{IEEEtran}
\IEEEoverridecommandlockouts

\usepackage{cite}
\usepackage{amsmath,amssymb,amsfonts}
\usepackage{algorithmic}
\usepackage{graphicx}
\usepackage{textcomp}
\usepackage{xcolor}
\usepackage{hyperref}
\usepackage{booktabs} 
\usepackage{colortbl} 
\usepackage{multirow} 

\def\BibTeX{{\rm B\kern-.05em{\sc i\kern-.025em b}\kern-.08em
    T\kern-.1667em\lower.7ex\hbox{E}\kern-.125emX}}
\makeatletter
\newcommand{\linebreakand}{%
  \end{@IEEEauthorhalign}
  \hfill\mbox{}\par
  \mbox{}\hfill\begin{@IEEEauthorhalign}
}
\makeatother
\begin{document}
\title{A Full-History Network Dataset for \\BTC Asset Decentralization Profiling\\
\thanks{*: Equal contribution, \dag: Corresponding author}
}

\author{\IEEEauthorblockN{1\textsuperscript{st} Ling Cheng*}
\IEEEauthorblockA{
\textit{Singapore Management University}\\
Singapore\\
lingcheng.2020@phdcs.smu.edu.sg}
\and
\IEEEauthorblockN{2\textsuperscript{nd} Qian Shao*}
\IEEEauthorblockA{
\textit{Singapore Management University}\\
Singapore\\
qianshao.2020@phdcs.smu.edu.sg}
\linebreakand 
\IEEEauthorblockN{3\textsuperscript{rd} Fengzhu Zeng}
\IEEEauthorblockA{
\textit{Singapore Management University}\\
Singapore\\
fzzeng.2020@phdcs.smu.edu.sg}
\and
\IEEEauthorblockN{4\textsuperscript{th} Feida Zhu\textsuperscript{\dag}}
\IEEEauthorblockA{
\textit{Singapore Management University}\\
Singapore\\
fdzhu@smu.edu.sg}
}
\maketitle

\begin{abstract}
Since its advent in 2009, Bitcoin (BTC) has garnered increasing attention from both academia and industry. 
However, due to the massive transaction volume, no systematic study has quantitatively measured the asset decentralization degree specifically from a network perspective.

In this paper, by conducting a thorough analysis of the BTC transaction network, we first address the significant gap in the availability of full-history BTC graph and network property dataset, which spans over 15 years from the genesis block (1st March, 2009) to the 845651-th block (29, May 2024). We then present the first systematic investigation to profile BTC's asset decentralization and design several decentralization degrees for quantification. Through extensive experiments, we emphasize the significant role of network properties and our network-based decentralization degree in enhancing Bitcoin analysis. Our findings demonstrate the importance of our comprehensive dataset and analysis in advancing research on Bitcoin's transaction dynamics and decentralization, providing valuable insights into the network's structure and its implications. The whole transaction data is available at \href{https://smu-my.sharepoint.com/:f:/g/personal/lingcheng_2020_phdcs_smu_edu_sg/EmebOEJ8i_RHnEJ3VOBjoUgBexiuMmJGs36-6XxA_bN1yg?e=TqdN5f}{dataset link}.
\end{abstract}


\section{Introduction}
\label{sec:intro}

Bitcoin (BTC)~\cite{16_} has transformed from a cryptography experiment into a global distributed system, trading at 70,000 USD as of June 4, 2024. As a digital currency enabling peer-to-peer transactions via cryptography and blockchain, BTC is often viewed as the prime application of distributed ledger technology and a symbol of decentralization. Despite frequent discussions on decentralization~\cite{5_, 7_, 8_, 9_}, limited research has assessed BTC’s decentralization as a financial asset. This perspective is crucial, as BTC is increasingly regarded as an asset, yet the relationship between its decentralization and financial phenomena remains unclear.

We present a full-history BTC transaction network dataset, spanning from the genesis block (March 1, 2009) to block 845,651 (May 2024), to systematically examine BTC's decentralization. Our analysis includes calculating key network properties for all addresses, tracking daily balances, identifying the top-5000 addresses by balance, and conducting community detection for cluster-level decentralization profiling. Focusing primarily on address-level analysis, we gain insights into BTC asset flow and distribution, complemented by cluster-level analysis through community detection. However, partial labeling of certain wallets and reliance on heuristic clustering algorithms may introduce biases, as many addresses lack wallet labels. To address this limitation, we plan to extend our work to the wallet level in future studies, aiming for more accurate clustering that can better distinguish true decentralization across individual and multi-address entities.

To profile BTC’s decentralization from a network perspective, we then analyze both the distribution of BTC tokens at regular snapshots and the flow of tokens through its transaction network. These metrics reflect the degree of decentralization of BTC in the context of a financial asset, offering insights into the network's structure and its implications for BTC as a decentralized financial asset. Through extensive experiments, including transaction fee and MVRV-Z score prediction (the market value's relative position to realized value), we emphasize the significant role of network properties and network-based decentralization in enhancing BTC analysis.


While specific to Bitcoin, this dataset and framework for measuring decentralization through network metrics and asset distribution are adaptable to other decentralized assets, enabling cross-chain comparisons. This scalability supports broader studies across blockchain networks and diverse applications. Our contributions can be summarized as follows:


\begin{itemize}
    \item We introduce the first full-history BTC transaction network dataset, covering from the genesis block to block 845651 (May 29, 2024). This dataset includes transaction networks, detailed network properties, top-balance addresses, and token distribution, serving as a valuable resource for BTC research and historical analysis.
    
    \item Using this dataset, we perform a longitudinal decentralization analysis, documenting key network properties and tracking top-balance addresses to reveal potential communities. Unlike prior snapshot-based studies, our approach identifies distinct phases in Bitcoin’s decentralization over time.
    
    \item Our experiments show that combining network properties with decentralization metrics improves BTC analysis tasks, such as transaction fee and MVRV-Z score prediction. This forecasting approach demonstrates the predictive value of decentralization metrics for financial analysis, offering new insights into blockchain research and network analysis.
\end{itemize}

The rest of the paper is organized as follows. We first describe our BTC transaction data in Section \ref{sec:dataset_detail}. 
In Section \ref{sec:decen_network_property}, \ref{sec:decen_asset_distribution}, \ref{sec:decen_rank_stability}, and \ref{sec:decen_hhi}, we conduct systematic analysis of various properties from both network and asset distribution. Several decentralization degrees are proposed to describe the decentralization from certain perspectives.
In Section \ref{sec:5_Downstream_Applications}, we conduct downstream applications to justify the effectiveness of network properties and our proposed metrics in the tasks related to BTC financial activities.
We then discuss related work in Section \ref{sec:6_Related_Work} and conclude the paper in Section \ref{sec:7_Conclusion}. 


\section{Dataset Details}
\label{sec:dataset_detail}

\subsection{Background}
\noindent \textbf{Blockchain and BTC:}
Blockchain is a decentralized ledger technology that securely records transactions across multiple computers, ensuring transparency and immutability. Each block in the chain contains a cryptographic hash of the previous block, forming a secure, tamper-resistant record of transactions. BTC is the first and most widely recognized cryptocurrency built on blockchain technology. It enables peer-to-peer transactions without the need for intermediaries, and each transaction is recorded on the BTC blockchain, with new blocks added through a process called mining, which requires computational power and energy consumption.

\noindent \textbf{UTXO Model:}
The BTC transaction system operates on the Unspent Transaction Output (UTXO) model, which differs significantly from the account-based model. Each BTC transaction consists of inputs and outputs, where inputs reference previous transaction outputs, and outputs define new recipients of BTC. Essentially, a BTC transaction consumes UTXOs and creates new ones. The sum of the inputs must equal or exceed the sum of the outputs, with any difference constituting a transaction fee collected by the miner who records the transaction in a block. Each transaction records every input and output, akin to double-entry bookkeeping. For example, in a transaction with a total input of 0.55 BTC and a total output of 0.50 BTC, the difference of 0.05 BTC is the miner's fee.

%



\subsection{Raw Data}
Our dataset spans from January 4, 2009, to May 29, 2024, covering more than 15 years of BTC transactions. Using BTC Core to sync with the full blockchain, we extracted all blocks from block-0 (Genesis) to block-845,652 with BlockSci~\cite{15_}, capturing transaction records, timestamps, confirmations, and fees. To further enrich the dataset, we calculated various metrics such as the total number of BTCs transacted, the average, maximum, and median transaction values as shown in Table.~\ref{tab:btc_data_summary}. In total, our dataset includes 2,593,912,022 transaction inputs and 2,846,704,728 outputs. This comprehensive dataset serves as a valuable resource for analyzing the historical and structural dynamics of the BTC network, enabling various research within the BTC ecosystem.

\begin{table}[h!]
\centering
\fontsize{10}{14}
\vspace{-1ex}
\caption{Summary of BTC Network Data}
\begin{tabular}{c|c}
\hline
\textbf{Description} & \textbf{Statistics} \\ \hline
Start date (mm-dd-yyyy, UTC)      & 01-03-2009 18:15 \\ \hline
End date (mm-dd-yyyy, UTC)        & 05-29-2024 14:24 \\ \hline
Number of unique addresses        & 1,411,482,182 \\ \hline
Number of transactions            & 1,013,989,952 \\ \hline
Number of blocks                  & 845,652 \\ \hline
Number of inputs                  & 2,593,912,022 \\ \hline
Number of outputs                 & 2,846,704,728 \\ \hline
Total BTC transacted (BTC)    & 8,654,507,676.117847 \\ \hline
Average transaction value (BTC)   & 8.535102008701015 \\ \hline
Maximum transaction value (BTC)   & 550,000.0 \\ \hline
Median transaction value (BTC)    & 0.03608083 \\ \hline
\end{tabular}
\label{tab:btc_data_summary}
\end{table}

\begin{figure*}
	\centering
	\vspace{-0ex}
	\includegraphics[width=2.0\columnwidth, angle=0]{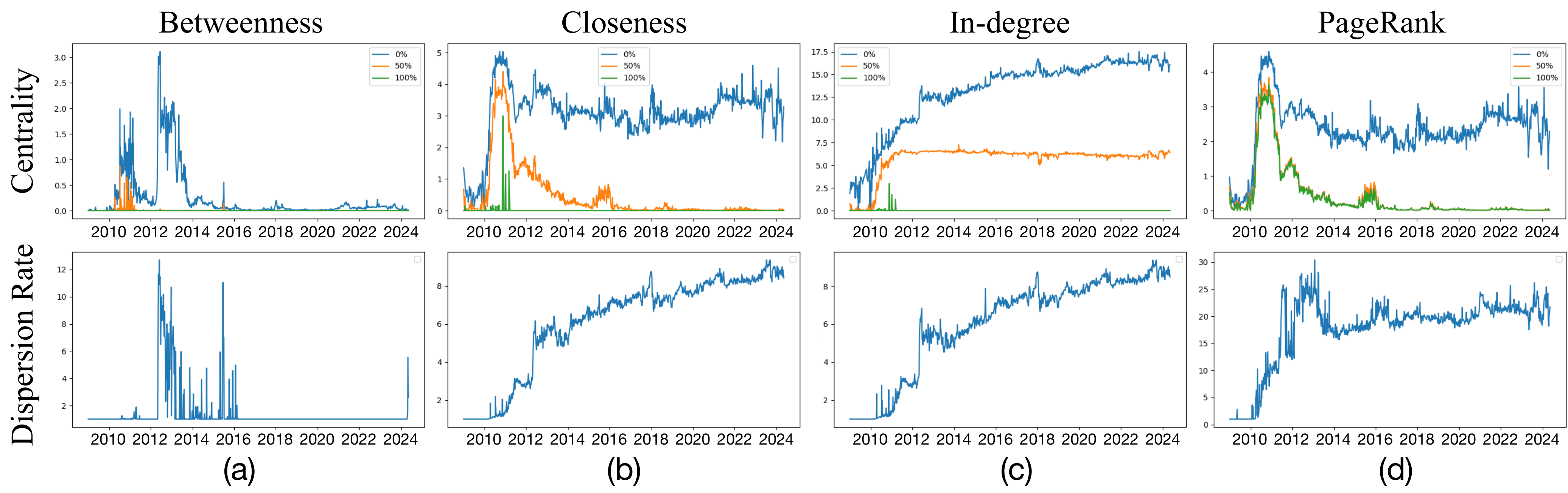}
 \vspace{-2ex}
\caption{Evolution of centrality and dispersion rate of four key centrality metrics: (a) Betweenness, (b) Closeness, (c) In-Degree, and (d) PageRank. These metrics are plotted for different percentiles of nodes. For dispersion rate, lower dispersion values suggest a more decentralized network structure.}
	\label{fig:Network_property_evo}
  \vspace{-2ex}
\end{figure*}
When BTC transactions are represented as a network, it becomes crucial to explore how network properties can measure dynamic decentralization. Graph centrality, which assesses the importance or influence of individual nodes, is naturally adopted for this analysis. We track the evolution of four key centrality metrics: Betweenness, Closeness, In-Degree, and PageRank to gain insights into the shifting dynamics of influence within the network.

\subsection{Graph Construction}
We processed the extracted blocks to build a detailed transaction dataset. Each transaction is broken down into its inputs and outputs, which represent the flow of BTCs from one address to another. 
However, in a BTC transaction, the input-output address mappings are not explicitly recorded, necessitating inference of edges between addresses based on transaction data. For a transaction with inputs from \( N \) distinct addresses and outputs to \( M \) distinct addresses, the following steps are followed:

\begin{enumerate}
    \item For each transaction, identify the set of input addresses and output addresses. Let \( I = \{i_1, i_2, \ldots, i_N\} \) represent the set of input addresses, and \( O = \{o_1, o_2, \ldots, o_M\} \) represent the set of output addresses.
    \item Create a directed edge from each input address to each output address, resulting in \( N \times M \) directed edges for the transaction~\cite{24_, 10_}. Each directed edge \( (i_k \rightarrow o_l) \) represents a potential transfer of BTCs from input address \( i_k \) to output address \( o_l \).
    \item Assign a weight to each edge based on the amount of BTCs transferred. If the transaction amount is not evenly distributed among outputs, weights are adjusted proportionally based on the output values.
\end{enumerate}
For example, consider a transaction with two input addresses \( A \) and \( B \) and three output addresses \( X \), \( Y \), and \( Z \). This transaction can be represented as:

\[
T = \{ (A, B) \rightarrow (X, Y, Z) \}.
\]
This transaction will be processed into the following directed edges, each edge \( (i_k \rightarrow o_l) \) can be weighted based on the transaction amount associated with the respective output address:
$\{A \rightarrow X, A \rightarrow Y, A \rightarrow Z, B \rightarrow X, B \rightarrow Y, B \rightarrow Z\}$.
%

\section{Decentralization from Network Properties}
\label{sec:decen_network_property}

\subsection{Evolution of Network Centrality}

\noindent \textbf{Betweenness}. Betweenness measures how often a node serves as a bridge in the shortest paths between other nodes. In the BTC transaction network, addresses with high betweenness centrality are key intermediaries that can influence the flow of BTC assets. Fig.~\ref{fig:Network_property_evo}(a) shows the evolution of Betweenness Centrality, which reflects a node's role as an intermediary in transactions. The peak around 2011-2012 suggests a period of centralization, likely linked to the rise of Mt. Gox. After this peak, Betweenness declines, indicating increasing decentralization as influence disperses across more nodes. We selected betweenness because it quantifies the influence of specific nodes (wallets) that frequently act as intermediaries in transaction paths. High betweenness values identify nodes critical to asset flow, reflecting potential centralization points where influence over transactions is concentrated. Analyzing betweenness evolution helps us understand shifts in control over asset flow as BTC’s ecosystem evolved.

\noindent \textbf{Closeness}. Closeness measures essentially the efficiency for spreading information across a network starting from a target node. In the BTC transaction network, addresses with higher closeness centrality tend to have a greater chance to impact other nodes' BTC asset through transaction relay. Fig.~\ref{fig:Network_property_evo}(b) illustrates the evolution of Closeness Centrality, which measures how quickly a node can interact with others. The data shows an initial rise in Closeness until around 2012, followed by stabilization, indicating that the network reached a point of equilibrium with no major shifts in node centrality. Closeness is crucial for capturing the speed and reach of asset movements, which is vital for understanding decentralization from a transactional accessibility standpoint.

\noindent \textbf{In-Degree}. In-degree represents asset transfers into a node, indicating trust and endorsement. The evolution of In-Degree Centrality, as shown in Fig.~\ref{fig:Network_property_evo}(c), reflects a growing concentration of incoming connections to a select group of nodes, signaling centralization trends. By capturing the volume of incoming connections, in-degree centrality highlights influential nodes, providing insight into shifts toward centralization.

\noindent \textbf{PageRank}. Transactions act as trust endorsements, with a higher PageRank score indicating greater trust~\cite{33_, 34_, 35_}. Fig.~\ref{fig:Network_property_evo}(d) shows PageRank Centrality's evolution, where early peaks suggest a few highly interconnected nodes held significant influence. PageRank, by weighing both quantity and quality of connections, highlights influential nodes and offers insights into asset centralization.

Overall, these centrality metrics highlight distinct phases in the BTC network’s evolution. Initially, certain nodes had significant influence, but as the network matured, there was a shift towards decentralization, with influence spreading across a broader array of nodes. This trend towards a more distributed network is critical for understanding BTC’s evolution and its broader implications for the cryptocurrency ecosystem.

\subsection{Dispersion Rate of Network Centrality}
\label{subsec:decen_network}

Decentralization in the BTC transaction network is not merely about centrality metrics but how these metrics are distributed among nodes, reflecting the degree of decentralization. To capture this, we introduce the concept of \emph{dispersion}, which quantifies how spread out or concentrated a metric is across the network. Dispersion for a metric $m$ is defined as:
\[
d_m = \log_2\left(1 + \frac{H_m - L_m}{M_m - L_m}\right),
\]
where $H_m$, $L_m$, and $M_m$ are the highest, lowest, and median values of the metric $m$, respectively.

Fig~\ref{fig:Network_property_evo} reveal that the BTC transaction network experienced phases of centralization and decentralization. The dispersion of \textbf{Betweenness} shows early fluctuations, indicating centralization around influential entities like Mt. Gox, followed by decentralization as influence spread. For \textbf{Closeness} and \textbf{In-Degree}, the dispersion rate steadily increased, reflecting a slow centralization process. The dispersion of \textbf{PageRank} peaked early, indicating initial high centralization, but later stabilized, showing a more balanced distribution of influence.
Overall, dispersion analysis across these centrality metrics provides a nuanced understanding of the BTC network's decentralization. While early stages show great fluctuation of decentralization, a steady trend towards lower decentralization emerges.

\section{Decentralization from Asset Distribution}
\label{sec:decen_asset_distribution}
It is customary to measure the centralization of traditional financial assets by identifying the so-called list of wealthiest individuals, e.g., the world's 26 richest people own as much as the poorest $50\%$ of the world's entire population. Likewise, for Bitcoin, we conduct the asset distribution analysis of its degree of decentralization by investigating two major questions and tracking the answers throughout its entire history: (1) Does a relatively small percentage of nodes on this ranking list control a significant portion of the total BTC asset? (2) How do nodes in the top-ranking list change over time?

\begin{figure*}
	\centering
	\vspace{-0ex}
	\includegraphics[width=2.0\columnwidth, angle=0]{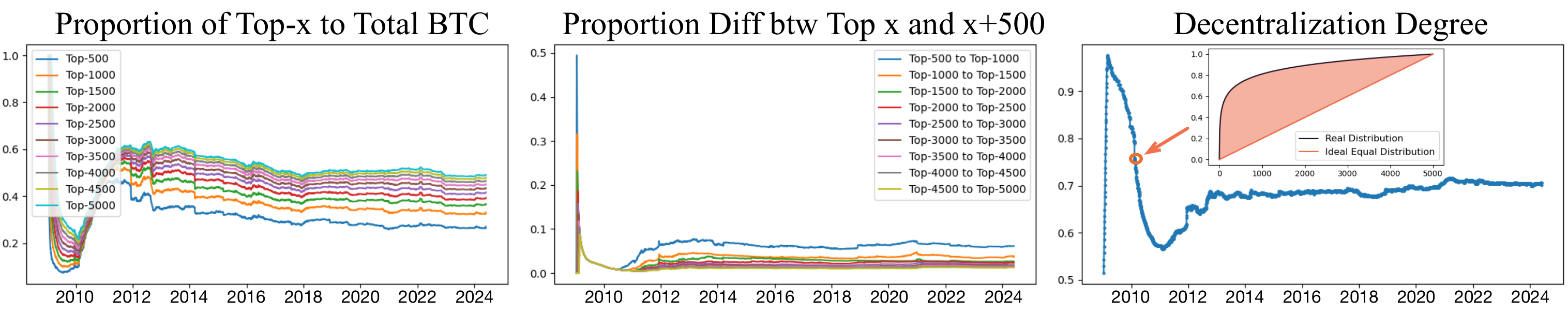}
 \vspace{-2ex}
    \caption{The left plot illustrates the proportion of total Bitcoins held by the top-$x$ addresses over time, with $x$ ranging from 500 to 5000 in increments of 500. The middle plot shows the proportional difference in Bitcoin holdings between consecutive top-$x$ groups, such as between the top-500 and top-1000, and so forth. The right plot displays the decentralization degree over time.}
	\label{fig:balance_proportion}
	\vspace{-0ex}
\end{figure*}

\subsection{Distribution Analysis}
\label{sec:static:proportion_evolution}
We first identify the top-5000 addresses by BTC balances on a daily basis and calculate the proportion of the total BTC asset commanded by the top-$x$ addresses for a range of values of $x$ from $500$ to $5000$ in increments of $500$ over time, as shown in Fig.~\ref{fig:balance_proportion}. Additionally, we zoom into each pair of top-$x$ and top-$(x+500)$ to calculate the difference in their BTC proportions among the total, and compare these differences across all pairs as $x$ ranges from $500$ to $4500$, as shown in the middle subgraph of Fig.~\ref{fig:balance_proportion}. The study presents that the entire Bitcoin history can be identified with three distinct phases based on two criteria, namely C1: the proportion owned by top-5000 addresses against the total BTC supply and C2: the proportional difference among different top-$x$ groups.

\vspace{2mm}
\noindent\textbf{Phase 1}: Increasing decentralization (Jan 9, 2009 to Feb 9, 2010). This phase is marked by significant decentralization as the Bitcoin network grows. The sharp decline in the proportion of Bitcoins held by the top-$x$ addresses in Fig.~\ref{fig:balance_proportion} reflects the influx of new participants and the wider distribution of BTC across more nodes. The dramatic decrease in the proportional differences between adjacent groups further confirms this rapid growth of decentralization.

\vspace{2mm}
\noindent\textbf{Phase 2}: Decreasing decentralization (Feb 10, 2010 to Mar 23, 2012). This phase witnesses a sharp decrease of decentralization both within the top-5000 addresses and against the total BTC asset, as shown by the simultaneous separating and rising of all the curves throughout this phase in Fig.~\ref{fig:balance_proportion}, serving as a clear sign for decreasing decentralization by both C1 and C2. 
The end of this phase is marked by a simultaneous climax for both C1 and C2, as illustrated by the peak and the maximum separation among all the curves, both of which are followed by a gradual decline heralding the beginning of Phase 3. 

\vspace{2mm}
\noindent\textbf{Phase 3}: Long-Term Stability with Minor Fluctuations (Mar 24, 2012 to present). The final phase is characterized by a stable distribution of BTC ownership among the top addresses, with minor fluctuations over time. As seen in Fig.~ \ref{fig:balance_proportion}, the curves flatten out, signifying that the top-500 to top-5000 addresses have maintained a relatively consistent share of the total BTC. The minimal differences between adjacent groups suggest a uniform distribution within these ranks, with only slight variations over time.

\subsection{Decentralization Degree on Asset Distribution}
After examining the top-ranking lists, it is crucial to study the overall decentralization degree of all these addresses, which we refer to as the \emph{Decentralization Degree on Asset Distribution}. In this analysis, we focus on the top 5000 addresses as they represent the majority of assets and activity within the BTC ecosystem.

The static decentralization degree is based on the concept that maximum decentralization occurs when all addresses hold an equal amount of BTC. In an ideal scenario, this would be depicted by a straight line on a cumulative distribution plot, where the cumulative proportion of BTCs owned by the top-$N$ addresses increases linearly with $N$. The actual distribution, however, forms a convex curve that deviates upward from this ideal line, with the extent of this deviation indicating the degree of centralization. The more pronounced the curve's bulge, the more uneven the BTC distribution among the top 5000 addresses, leading to lower decentralization.

We define this decentralization degree, denoted by \(D_{A}\), by measuring the area between the actual cumulative distribution curve and the ideal equal distribution line. Specifically, \(D_{A}\) is calculated as:
\[
D_{A} = 1-\int_0^{N} \left(\mathbb{C}_r(x) - \mathbb{C}_e(x)\right) dx,
\]
where \(\mathbb{C}_r(x)\) and \(\mathbb{C}_e(x)\) represent the cumulative proportion of BTCs owned by the top-$x$ addresses in the real distribution and the ideal distribution, respectively. The rightmost subplot in Fig.~\ref{fig:balance_proportion} illustrates this decentralization degree. We can have a clearer overview of the decentralization degree for the full BTC history. The distinct patterns exhibited by different segments of the curve align well with the boundaries of the three phases, a strong support for our definition, and characterizations of decentralization degree, of the three phases from another perspective.

\section{Decentralization on Ranking Stability}
\label{sec:decen_rank_stability}
\begin{figure*}
	\centering
	\vspace{-0ex}
	\includegraphics[width=2.0\columnwidth, angle=0]{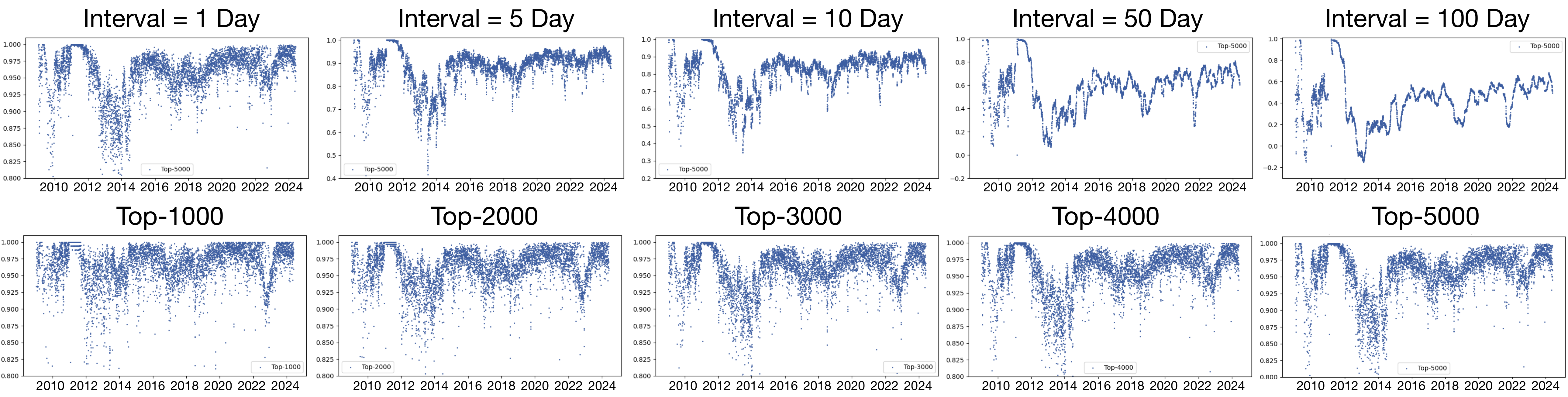}
  \vspace{-2ex}
\caption{Spearman coefficient analysis of BTC top-ranking addresses. (1) Spearman coefficients for different day intervals for top-5000 addresses. (2) Breakdown of Spearman coefficients across different ranking groups with 1 Day interval. A higher Spearman coefficient indicates greater stability.}
	\label{fig:spearman_analysis}
	\vspace{-0ex}
\end{figure*}

\begin{figure*}
	\centering
	\vspace{-0ex}
	\includegraphics[width=2.0\columnwidth, angle=0]{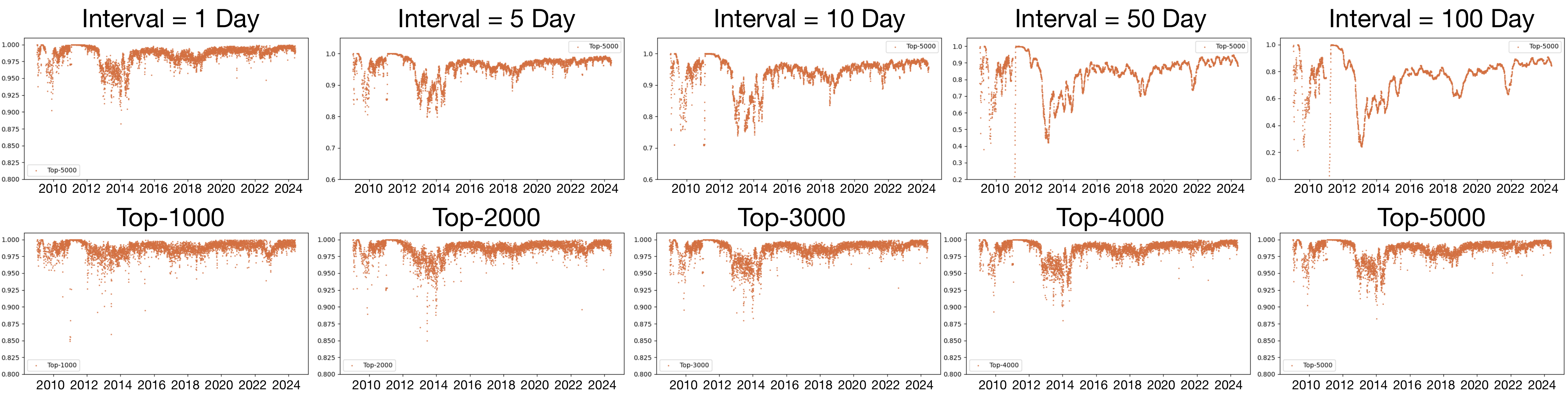}
 \vspace{-2ex}
\caption{Retention rate analysis of BTC top-ranking addresses: (1) Retention rates for different day intervals for top-5000 addresses. (2) Breakdown of retention rates across different ranking groups with 1 Day interval. A higher retention rate indicates a greater proportion of addresses remain.}
	\label{fig:retention_analysis}
	\vspace{-2ex}
\end{figure*}

Another important aspect of decentralization is the evaluation of the stability of the membership and order within the top-ranking lists over time, which we refer to as \emph{ranking stability}. The more stable the ranking, the lower the degree of decentralization. To assess how the top-ranking list evolves, we use two complementary measures: (I) Spearman coefficient and (II) Retention rate. 

The Spearman coefficient (variables correlation metric) is employed to assess how the order of nodes within the ranking changes over time as an indicator of ranking stability. For two rankings on day \(i\) and day \(i+n\) (an \(n\)-day interval), we take the ranking on day \(i\) as variable \(X\) and the ranking on day \(i+n\) as \(Y\), and calculate the Spearman coefficient for \(X\) and \(Y\). Higher absolute value of Spearman coefficient indicates high consistency between the two rankings, reflecting high ranking stability. The Retention rate is used to measure the extent of membership change in the top-$N$ lists. It is calculated as the ratio of the number of retained addresses to \(N\), the size of the ranking list. These two measures complement each other to provide a comprehensive evaluation of ranking stability; for example, a ranking list might show stability in membership but still experience significant changes in the relative order of its members.
The analysis of ranking stability reveals several key findings:

\noindent \textbf{From Interval:} As the interval increases, the composition of the top-5000 addresses experiences greater changes. However, over time, the overall trend tends to stabilize, aligning with the previously observed trend in the decentralization degree. A significant reduction is observed between 2011 and 2012, followed by a gradual increase. Although there are noticeable fluctuations throughout the process, the overall trend of increase remains unchanged. 

By comparing the Spearman coefficient and retention rate, it is evident that when the interval is set to 1 to 10 days, the retention rate shows greater variation, indicating that short-term changes are mainly due to the addition of new addresses, with relatively minor changes in the balance ranking. In the long term, with intervals of 50 and 100 days, the Spearman coefficient shows greater variability, suggesting that over the long term, the proportion of new address additions remains stable, while significant changes occur in the balance ranking of the top-5000 addresses.

\noindent \textbf{From Group Size:} The ranking stability is higher for groups positioned closer to the top, as reflected by higher Spearman coefficients and retention rates compared to other groups. As the value of top-x increases, the changes within the group become more consistent, leading to a decrease in variance for both the Spearman coefficient and retention rate. This suggests that the top-5000 addresses likely hold fixed roles, such as exchanges or mining pools. These addresses with fixed identities tend to exhibit uniform responses to network events, which contributes to the observed reduction in variance.

\section{Decentralization on Market Efficiency}
\label{sec:decen_hhi}
BTC transactions can be viewed as a financial market, making it useful to study the network's decentralization from a market efficiency perspective. The Herfindahl-Hirschman Index (HHI) is a classic measure of market concentration~\cite{36_, 37_}, defined as:
\begin{equation}
\label{eq:HHI_calculation}
HHI = \sum_{i=1}^{n} 10000 \times (H_i)^2, \hspace{1mm} H_i = \frac{h_i}{C},
\end{equation}
where $h_i$ is the BTC held by entity $i$, and $C$ is the total BTC minted by that time. A higher HHI indicates more market concentration and less decentralization.
In our context, the concept of a "firm" in the original HHI definition is interpreted as an entity controlling BTC. Due to BTC's anonymity, it's challenging to identify if a set of addresses belongs to the same entity. We propose two approaches: (1) Treat each address as a distinct firm, or (2) Cluster addresses into entities using community detection on a full-history transaction graph of top-5000 addresses, with all communities treated as separate firms.
We thus propose a decentralization degree $D_{HHI}$ based on the A1 and A2 clustering approach, A higher $D_{HHI}$ indicates greater decentralization.
\begin{equation}
\label{eq:metric_2}
D_{HHI} = 1 - \text{normalized}(HHI_{A1/A2}).
\end{equation}

\begin{figure}
    \vspace{-0ex}
    \includegraphics[width=1.\columnwidth, angle=0, scale=1.0]{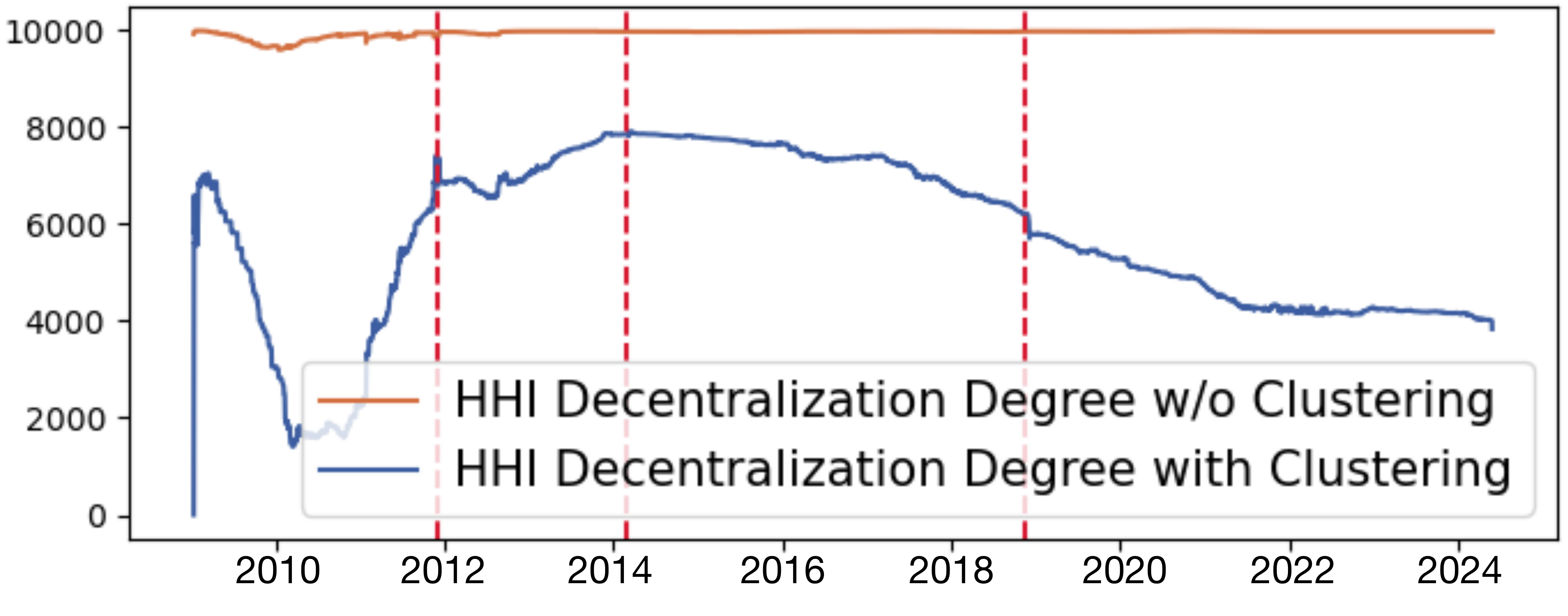}
    \vspace{-5ex}
\caption{HHI decentralization degree, with independent entities (orange) and clustered entities (black). Red dashed lines highlight key dates impacting BTC decentralization.}
    \label{fig:HHI_analysis}
    \vspace{-2ex}
\end{figure}

Fig.~\ref{fig:HHI_analysis} presents the HHI decentralization degree calculated for all top-5000 addresses throughout BTC history. Different colored curves correspond to the two clustering approaches. Based on the result, we have the following findings:
\begin{itemize}
\item A1 shows a consistently lower HHI, indicating that no single address has dominated the market, which suggests a high level of decentralization when treating all addresses as independent entities. This also implies that significant entities tend to control multiple addresses rather than concentrating their holdings in a single address.
\item A2 exhibits a sharp rise early in BTC history as more participants joined the network. With the entry of large institutions, a substantial amount of capital was concentrated in the hands of these major players, leading to a sharp increase in HHI. However, considering the events involving large exchanges between 2010 and 2011, tokens became more dispersed among a broader range of entities, resulting in a subsequent decrease in HHI.
\end{itemize}
We also highlight three significant events corresponding to major changes in $D_{HHI}$:

\vspace{2mm}
\noindent\textbf{(1) December 7, 2011:} After BTC's first major bubble burst, prices stabilized around \$2-\$3, reflecting recovery and growing user acceptance despite earlier volatility.

\vspace{2mm}
\noindent\textbf{(2) March 16, 2014:} The collapse of Mt. Gox led to significant market fluctuations and a loss of confidence in exchanges, impacting decentralization.

\vspace{2mm}
\noindent\textbf{(3) December 5, 2018:} A drop in BTC mining difficulty raised concerns about a 51\% attack, triggering panic selling and an abrupt change in $D_{HHI}$.

\section{Downstream Applications}
\label{sec:5_Downstream_Applications}
In this section, we conduct comprehensive experiments to examine how the integration of network properties and our decentralization metrics impacts the performance of BTC analysis tasks, including transaction fee forecasting and MVRV(Market Value to Realized Value)-Z score forecasting. Transaction fee forecasting is crucial for optimizing user costs and improving network efficiency, while providing insights into Bitcoin network activity. MVRV-Z score forecasting helps investors assess whether Bitcoin is overvalued or undervalued, aiding in market timing and risk management. Precise prediction of these metrics can offer deeper insights into BTC's dynamics.

To tackle these forecasting tasks, we define them as time series forecasting problems. Specifically, the time series forecasting problem is defined as follows: given historical observations $\mathbf{X} = \{ \mathbf{x}_1, \dots, \mathbf{x}_T \} \in \mathbb{R}^{T \times N}$ with $T$ time steps and $N$ variables, we aim to predict the next $S$ time steps $\mathbf{Y} = \{ \mathbf{y}_{T+1}, \dots, \mathbf{y}_{T+S} \} \in \mathbb{R}^{S \times 1}$. The $N$ variables, denoted as $\mathbf{x}_t$ at time $t$, include \textit{Centrality}, \textit{Asset Distribution}, \textit{Rank Stability} features, along with transaction fee or MVRV-Z. The target variable $\mathbf{y}_t$ corresponds to either the transaction fee or MVRV-Z for the given prediction horizon.

Since \textit{Centrality} features are calculated at the block level, we aggregate them to the daily level by computing the average, minimum, and maximum values for each day. The dataset is split into training, validation, and test sets with a 70/10/20 ratio. We use the following state-of-the-art and traditional baselines:

\begin{itemize}
    \item \textbf{iTransformer\cite{liu2023itransformer}} repurposes the Transformer architecture by applying attention and feed-forward networks to inverted dimensions, enabling improved multivariate correlations and performance on time series forecasting with arbitrary lookback windows.
    \item \textbf{Flashformer\cite{dao2022flashattention}} refers to a Transformer equipped with hardware-accelerated FlashAttention, an IO-aware attention algorithm that optimizes memory access to significantly accelerate training on long sequences.
    \item \textbf{Flowformer\cite{wu2022flowformer}} utilizes Flow-Attention, a linear-time attention mechanism based on flow network theory, enabling efficient and bias-free processing of long sequences across various domains.
    \item \textbf{Reformer\cite{kitaev2019reformer}} is Transformer-based model that reduces attention complexity with locality-sensitive hashing and reversible residual layers for long sequences processing.
    \item \textbf{Transformer\cite{vaswani2017attention}} is solely on attention mechanisms, offering superior performance and faster training in sequence transduction tasks without the need for recurrence.
    \item \textbf{GRU\cite{cho2014learning}} is a typical neural network for sequence modeling. At each time split, previous hidden state and current summation features are fed into the GRU unit to predict the labels for the given addresses
    \item \textbf{LSTM \cite{hochreiter1997long}} is a type of recurrent neural network (RNN) designed to capture long-term dependencies in sequential data, using memory cells and gates to overcome the vanishing gradient problem.   
\end{itemize}

The performance of the models is assessed using Mean Squared Error (MSE) and Mean Absolute Error (MAE), The lower MSE and MAE indicate the more accurate prediction result. We aim answer the following research questions:

\textbf{Q1}: How does the length of historical data, in combination with varying prediction horizons, impact the accuracy of transaction fee and MVRV predictions across different models?

\textbf{Q2}: How do different combinations of \textit{Centrality}, \textit{Asset Distribution}, and \textit{Rank Stability} features influence prediction performance across various models?

\subsection{Impact of Historical and Prediction Horizons (Q1)}

\begin{table*}[h]
    \centering
    \caption{Performance comparison of transaction fee across different models and different prediction length(His. denotes historical length, and Pred. denotes prediction length). The best results are highlighted in bold.}
    \resizebox{\textwidth}{!}{
    \begin{tabular}{ll|cc|cc|cc|cc|cc|cc|cc|cc}
        \toprule
          \multicolumn{2}{c}{\textbf{Length}} & \multicolumn{2}{c}{\multirow{2}{*}{\vspace{1cm} \textbf{iTransformer} }} & \multicolumn{2}{c}{\multirow{2}{*}{\vspace{1cm}\textbf{Flashformer}}} & \multicolumn{2}{c}{\multirow{2}{*}{\vspace{1cm} \textbf{Flowformer}}} & \multicolumn{2}{c}{\multirow{2}{*}{\vspace{1cm} \textbf{Informer}}} & \multicolumn{2}{c}{\multirow{2}{*}{\vspace{1cm} \textbf{Reformer}}} & \multicolumn{2}{c}{\multirow{2}{*}{\vspace{1cm} \textbf{Transformer}}} & \multicolumn{2}{c}{\multirow{2}{*}{\vspace{1cm} \textbf{GRU}}} & \multicolumn{2}{c}{\multirow{2}{*}{\vspace{1cm} \textbf{LSTM}}} \\
        \cmidrule(lr){3-4} \cmidrule(lr){5-6} \cmidrule(lr){7-8} \cmidrule(lr){9-10} \cmidrule(lr){11-12} \cmidrule(lr){13-14} \cmidrule(lr){15-16} \cmidrule(lr){17-18}
        \textbf{His.} & \textbf{Pred.} & \textbf{MSE} & \textbf{MAE} & \textbf{MSE} & \textbf{MAE} & \textbf{MSE} & \textbf{MAE} & \textbf{MSE} & \textbf{MAE} & \textbf{MSE} & \textbf{MAE} & \textbf{MSE} & \textbf{MAE} & \textbf{MSE} & \textbf{MAE} & \textbf{MSE} & \textbf{MAE} \\
        \midrule
       
        \textbf{30} & \textbf{1} & \textbf{0.190} & \textbf{0.135} & 0.794 & 0.762 & 0.674 & 0.686 & 0.915 & 0.810 & 0.647 & 0.597 & 0.635 & 0.647 & 0.651 & 0.654 & 1.155 & 0.902 \\

        \textbf{14} & \textbf{1} &\textbf{0.208} & \textbf{0.129} & 0.549 & 0.492 & 0.554 & 0.490 & 0.695 & 0.678 & 0.687 & 0.472 & 0.737 & 0.644  & 0.597 & 0.612 & 1.090 & 0.881 \\

        \textbf{7} &\textbf{1} & \textbf{0.208} & \textbf{0.123} & 0.587 & 0.601 & 0.541 & 0.585 & 0.586 & 0.601 & 0.698 & 0.634 & 0.705 & 0.688 & 0.530 & 0.580 & 0.946 & 0.832 \\

        \midrule
        \textbf{30} &\textbf{7} & \textbf{0.374} & \textbf{0.223} & 1.366 & 0.991 & 1.390 & 0.959 & 1.064 & 0.826 & 1.055 & 0.815 & 1.484 & 1.038 &   1.641 & 1.040 & 2.267 & 1.124  \\

        \textbf{14} &\textbf{7} & \textbf{0.347} & \textbf{0.196} & 1.577 & 1.074 & 1.572 & 1.023 & 1.643 & 1.040 & 1.206 & 0.882 & 1.682 & 1.119 & 1.766 & 1.082 & 1.839 & 1.049 \\

        \textbf{7} &\textbf{7} & \textbf{0.354} & \textbf{0.194} & 1.117 & 0.853 & 1.156 & 0.851 & 1.186 & 0.838 & 1.415 & 0.958 & 1.353 & 0.942 & 1.490 & 1.005 & 1.786 & 1.056 \\

        \midrule
        \textbf{30} &\textbf{14} & \textbf{0.420} & \textbf{0.244}& 1.444 & 0.959 & 1.646 & 1.043 & 1.131 & 0.727 & 1.023 & 0.746 & 1.629 & 1.018 & 2.235 & 1.195 & 2.247 & 1.195 \\

        \textbf{14} & \textbf{14} & \textbf{0.474} & \textbf{0.250} & 1.261 & 0.895 & 1.017 & 0.796 & 1.344 & 0.894 & 1.693 & 1.063 & 2.053 & 1.179 & 2.301 & 1.211 & 1.868 & 1.056 \\

        \textbf{30} & \textbf{30} & \textbf{0.582} & \textbf{0.311}& 1.696 & 0.937 & 1.489 & 0.858 & 0.880 & 0.640 & 0.721 & 0.565 & 1.806 & 0.973 & 3.526 & 1.400 & 1.734 & 0.927 \\
          
        \bottomrule
    \end{tabular}
    }
    \label{tab: length fee}
\end{table*}

\begin{figure*}
    \vspace{-0ex}
    \includegraphics[width=\textwidth]{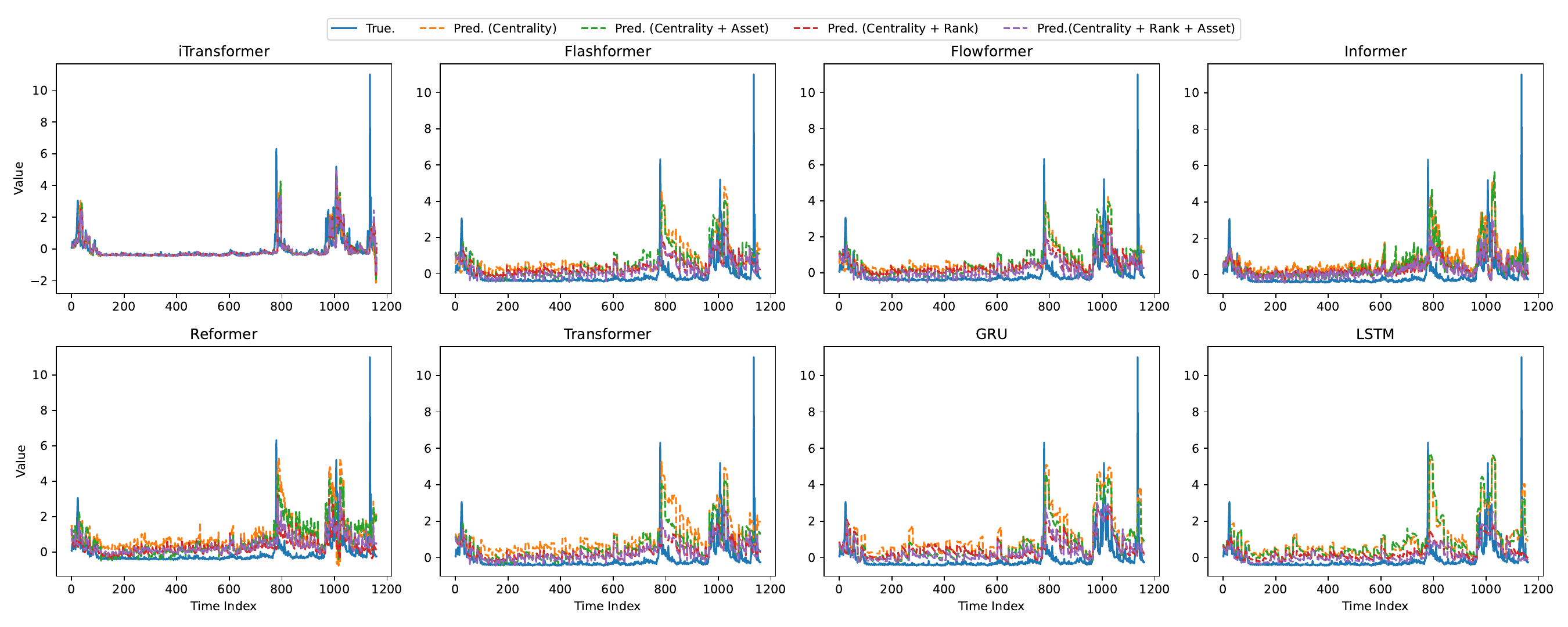}
    \vspace{-5ex}
\caption{Visualization of 14-day transaction fee predictions based on different combinations of features, using historical 14-day data.}
    \label{fig:compare fee}
    \vspace{-2ex}
\end{figure*}

\begin{table*}[h]
    \centering
    \caption{Performance comparison of transaction fee across models and feature combination. The best results are in bold.}
    \resizebox{\textwidth}{!}{
    \begin{tabular}{l|cc|cc|cc|cc|cc|cc|cc|cc}
        \toprule
         \textbf{Features} & \multicolumn{2}{c}{\multirow{2}{*}{\vspace{1cm} \textbf{iTransformer} }} & \multicolumn{2}{c}{\multirow{2}{*}{\vspace{1cm}\textbf{Flashformer}}} & \multicolumn{2}{c}{\multirow{2}{*}{\vspace{1cm} \textbf{Flowformer}}} & \multicolumn{2}{c}{\multirow{2}{*}{\vspace{1cm} \textbf{Informer}}} & \multicolumn{2}{c}{\multirow{2}{*}{\vspace{1cm} \textbf{Reformer}}} & \multicolumn{2}{c}{\multirow{2}{*}{\vspace{1cm} \textbf{Transformer}}} & \multicolumn{2}{c}{\multirow{2}{*}{\vspace{1cm} \textbf{GRU}}} & \multicolumn{2}{c}{\multirow{2}{*}{\vspace{1cm} \textbf{LSTM}}} \\
        
        \cmidrule(lr){2-3} \cmidrule(lr){4-5} \cmidrule(lr){6-7} \cmidrule(lr){8-9} \cmidrule(lr){10-11} \cmidrule(lr){12-13} \cmidrule(lr){14-15} \cmidrule(lr){16-17}
         & \textbf{MSE} & \textbf{MAE} & \textbf{MSE} & \textbf{MAE} & \textbf{MSE} & \textbf{MAE} & \textbf{MSE} & \textbf{MAE} & \textbf{MSE} & \textbf{MAE} & \textbf{MSE} & \textbf{MAE} & \textbf{MSE} & \textbf{MAE} & \textbf{MSE} & \textbf{MAE} \\
        \midrule
       
        \textbf{Centrality} & 0.474 & 0.250 & 1.261 & 0.895 & 1.017 & 0.796 & 1.344 & 0.894 & 1.693 & 1.063 & 2.053 & 1.179 & 2.301 & 1.211 & 1.868 & 1.056 \\

        \textbf{Centrality + Asset} & 0.456 & 0.244 &  1.115 & 0.791 & 1.049 & 0.763 & 0.987 & 0.691 & 1.304 & 0.866 & 1.154 & 0.810 & 1.434 & 0.891 & 2.023 & 1.054 \\

        \textbf{Centrality + Rank} & \textbf{0.435} & 0.238 & 0.630 & 0.599 & 0.642 & 0.615 & 0.639 & 0.586 & 0.610 & \textbf{0.547} & 0.621 & 0.586 & 0.825 & 0.708 & 0.498 & 0.504 \\

        \textbf{Centrality + Asset + Rank} & \textbf{0.435} & \textbf{0.237} & \textbf{0.559} & \textbf{0.509} & \textbf{0.525} & \textbf{0.471} & \textbf{0.637} & \textbf{0.551} & \textbf{0.585} & 0.578 & \textbf{0.590} & \textbf{0.549} & \textbf{0.671} & \textbf{0.588} & \textbf{0.465} & \textbf{0.416} \\

        \bottomrule
    \end{tabular}
    }
    \label{t3: fee feature}
\end{table*}





We investigate the impact of historical data lengths (30, 14, and 7 days) on prediction accuracy across various models for different prediction horizons (1, 7, 14, and 30 days). Tables~\ref{tab: length fee} and~\ref{tab: length mvrv} show model performance in predicting transaction fees and MVRV-Z using Centrality features across these lengths. For 1-day predictions, iTransformer consistently achieves the lowest errors across all historical lengths, while other models show higher errors. For 7-day predictions, errors increase but iTransformer remains the best, particularly for 7-to-7 predictions in transaction fees (MSE: 0.354, MAE: 0.194) and 30-to-7 in MVRV-Z (MSE: 0.010). Other models, especially GRU and LSTM, struggle with medium- and long-term predictions. Overall, iTransformer outperforms across all tasks, with its advantage growing as prediction horizons lengthen. GRU and LSTM perform acceptably for shorter predictions but show significant errors over longer horizons.

\subsection{Impact of Feature Combinations on Prediction Performance (Q2)}

\begin{figure*}
    \vspace{-0ex}
    \includegraphics[width=\textwidth]{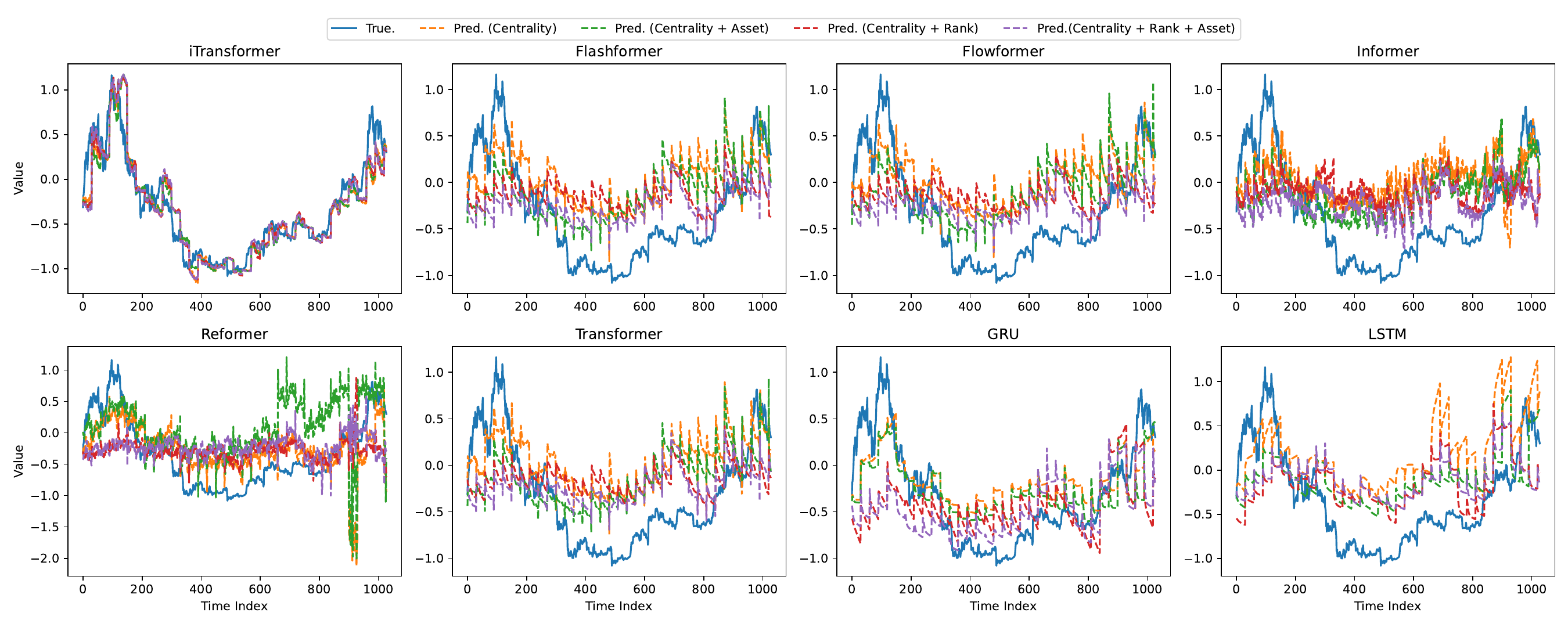}
    \vspace{-5ex}
\caption{Visualization of 30-day MVRV-Z predictions based on different combinations of features, using historical 30-day data.}
    \label{fig:compare MVRV}
    \vspace{-2ex}
\end{figure*}

\begin{table*}[h]
    \centering
    \caption{Performance comparison of MVRV-Z across different models and different prediction length(His. denotes historical length, and Pred. denotes prediction length). The best results are highlighted in bold.}
    \resizebox{\textwidth}{!}{
    \begin{tabular}{ll|cc|cc|cc|cc|cc|cc|cc|cc}
        \toprule
          \multicolumn{2}{c}{\textbf{Length}} & \multicolumn{2}{c}{\multirow{2}{*}{\vspace{1cm} \textbf{iTransformer} }} & \multicolumn{2}{c}{\multirow{2}{*}{\vspace{1cm}\textbf{Flashformer}}} & \multicolumn{2}{c}{\multirow{2}{*}{\vspace{1cm} \textbf{Flowformer}}} & \multicolumn{2}{c}{\multirow{2}{*}{\vspace{1cm} \textbf{Informer}}} & \multicolumn{2}{c}{\multirow{2}{*}{\vspace{1cm} \textbf{Reformer}}} & \multicolumn{2}{c}{\multirow{2}{*}{\vspace{1cm} \textbf{Transformer}}} & \multicolumn{2}{c}{\multirow{2}{*}{\vspace{1cm} \textbf{GRU}}} & \multicolumn{2}{c}{\multirow{2}{*}{\vspace{1cm} \textbf{LSTM}}} \\
        \cmidrule(lr){3-4} \cmidrule(lr){5-6} \cmidrule(lr){7-8} \cmidrule(lr){9-10} \cmidrule(lr){11-12} \cmidrule(lr){13-14} \cmidrule(lr){15-16} \cmidrule(lr){17-18}
        \textbf{His.} & \textbf{Pred.} & \textbf{MSE} & \textbf{MAE} & \textbf{MSE} & \textbf{MAE} & \textbf{MSE} & \textbf{MAE} & \textbf{MSE} & \textbf{MAE} & \textbf{MSE} & \textbf{MAE} & \textbf{MSE} & \textbf{MAE} & \textbf{MSE} & \textbf{MAE} & \textbf{MSE} & \textbf{MAE} \\
        \midrule
       
        \textbf{30} & \textbf{1} & \textbf{0.003} & \textbf{0.035} & 0.209 & 0.380 & 0.187 & 0.354 & 0.107 & 0.268 & 0.099 &0.232 & 0.111 & 0.283 & 0.049 & 0.184 & 0.142 & 0.321 \\

       \textbf{14} & \textbf{1} & \textbf{0.003} & \textbf{0.033} & 0.166 & 0.355 & 0.165 & 0.354 & 0.346 & 0.504 & 0.159 & 0.350 & 0.115 & 0.296 & 0.122 & 0.305 & 0.174 & 0.369 \\

        \textbf{7} & \textbf{1} & \textbf{0.003} & \textbf{0.032} & 0.086 & 0.257 & 0.087 & 0.258 & 0.113 & 0.292 & 0.117 & 0.305 & 0.090 & 0.262 & 0.103 & 0.278 & 0.186 & 0.385 \\

        \midrule

        \textbf{30} &\textbf{7}& \textbf{0.012} & \textbf{0.073} & 0.229 & 0.415 & 0.199 & 0.381 & 0.214 & 0.406 & 0.186 & 0.317 & 0.161 & 0.351 & 0.100 & 0.268 & 0.092 & 0.255 \\

       \textbf{14} &\textbf{7} & \textbf{0.009} & \textbf{0.064} & 0.175 & 0.361 & 0.166 & 0.353 & 0.199 & 0.381 & 0.200 & 0.382 & 0.171 & 0.357 & 0.163 & 0.344 & 0.202 & 0.392 \\

        \textbf{7} &\textbf{7} & \textbf{0.010} & \textbf{0.064} & 0.157 & 0.346 & 0.163 & 0.355 & 0.156 & 0.337 & 0.199 & 0.373 &0.146 & 0.331 & 0.121 & 0.293 & 0.288 & 0.480 \\
        
        \midrule

        \textbf{30} &\textbf{14} & \textbf{0.021} & \textbf{0.098} & 0.243 & 0.431 & 0.217 & 0.407 & 0.249 & 0.440 & 0.232 & 0.374 & 0.218 & 0.412 & 0.094 & 0.252 & 0.123 & 0.290 \\

        \textbf{14} & \textbf{14} & \textbf{0.016} & \textbf{0.085} & 0.210 & 0.397 & 0.206 & 0.396 & 0.221 & 0.405 & 0.241 & 0.410 & 0.204 & 0.389 & 0.175 & 0.333 & 0.279 & 0.470 \\

        \textbf{30} & \textbf{30} & \textbf{0.040 }& \textbf{0.137} & 0.279 & 0.469 & 0.275 & 0.464 & 0.356 & 0.530 & 0.252 & 0.384 & 0.283 & 0.470 & 0.149 & 0.327 & 0.470 & 0.606 \\

        \bottomrule
    \end{tabular}
    }
    \label{tab: length mvrv}
    \vspace{-1ex}

\end{table*}

\begin{table*}[h]
    \centering
    \caption{Performance comparison of MVRV-Z across models and feature combination. The best results in bold.}
    \resizebox{\textwidth}{!}{
    \begin{tabular}{l|cc|cc|cc|cc|cc|cc|cc|cc}
        \toprule
         \textbf{Features} & \multicolumn{2}{c}{\multirow{2}{*}{\vspace{1cm} \textbf{iTransformer} }} & \multicolumn{2}{c}{\multirow{2}{*}{\vspace{1cm}\textbf{Flashformer}}} & \multicolumn{2}{c}{\multirow{2}{*}{\vspace{1cm} \textbf{Flowformer}}} & \multicolumn{2}{c}{\multirow{2}{*}{\vspace{1cm} \textbf{Informer}}} & \multicolumn{2}{c}{\multirow{2}{*}{\vspace{1cm} \textbf{Reformer}}} & \multicolumn{2}{c}{\multirow{2}{*}{\vspace{1cm} \textbf{Transformer}}} & \multicolumn{2}{c}{\multirow{2}{*}{\vspace{1cm} \textbf{GRU}}} & \multicolumn{2}{c}{\multirow{2}{*}{\vspace{1cm} \textbf{LSTM}}} \\
        \cmidrule(lr){2-3} \cmidrule(lr){4-5} \cmidrule(lr){6-7} \cmidrule(lr){8-9} \cmidrule(lr){10-11} \cmidrule(lr){12-13} \cmidrule(lr){14-15} \cmidrule(lr){16-17}
         & \textbf{MSE} & \textbf{MAE} & \textbf{MSE} & \textbf{MAE} & \textbf{MSE} & \textbf{MAE} & \textbf{MSE} & \textbf{MAE} & \textbf{MSE} & \textbf{MAE} & \textbf{MSE} & \textbf{MAE} & \textbf{MSE} & \textbf{MAE} & \textbf{MSE} & \textbf{MAE} \\
        \midrule

        \textbf{Centrality} & 0.040 & 0.137 & 0.279 & 0.469 & 0.275 & 0.464 & 0.356 & 0.530 & \textbf{0.252} & \textbf{0.384 }& 0.283 & 0.470 & 0.149 & 0.327 & 0.470 & 0.606 \\

        \textbf{Centrality + Asset} & 0.041 & 0.137 & 0.268 & 0.449 & \textbf{0.252} & \textbf{0.434} &\textbf{ 0.262} & \textbf{0.446} & 0.437 & 0.568 & \textbf{0.257 }& \textbf{0.440} & \textbf{0.127} & \textbf{0.292} & \textbf{0.221} & \textbf{0.412} \\

        \textbf{Centrality + Rank} & 0.040 & 0.138 & 0.302 & 0.477 & 0.303 & 0.481 & 0.323 & 0.497 & 0.289 & 0.449 & 0.297 & 0.471 & 0.261 & 0.380  & 0.302 & 0.475 \\

        \textbf{Centrality + Asset + Rank} & \textbf{0.039} & \textbf{0.136} & \textbf{0.260} & \textbf{0.432} & 0.264 & 0.436 & 0.294 & 0.466 & 0.308 & 0.472 & 0.269 & 0.444 & 0.251 & 0.366 & 0.289 & 0.478 \\

        \bottomrule
    \end{tabular}
    }
    \label{t4: mvrv feature}
    \vspace{-1.5ex}
\end{table*}

While the length of historical data is crucial, the choice of features also plays a significant role in prediction accuracy. In this section, we explore how different combinations of \textit{Centrality} features, \textit{Asset Distribution} features, and \textit{Rank Stability }features impact the performance of transaction fee prediction and MVRV-Z prediction. 

The addition of \textit{Asset} and \textit{Rank} features improves the performance of models in both the transaction fee prediction (Table.~\ref{t3: fee feature}) and MVRV-Z prediction (Table.~\ref{t4: mvrv feature}) tasks, compared to using only \textit{Centrality} features. This is further supported by visualizations in Fig.~\ref{fig:compare fee} and Fig.~\ref{fig:compare MVRV}, which show how different feature combinations affect predictions.

For the transaction fee prediction (Table.~\ref{t3: fee feature}  and Fig.~\ref{fig:compare fee}), adding \textit{Asset} to \textit{Centrality} reduces the MSE for iTransformer from 0.474 to 0.456, and further decreases to 0.435 when \textit{Rank} is also added. A similar trend is observed for Flowformer and Reformer, where the full combination of features (\textit{Centrality} + \textit{Asset} + \textit{Rank}) yields the lowest error values. In Fig.~\ref{fig:compare fee} , predictions using all features align more closely with the true values (blue line). For the MVRV-Z prediction task (Table.~\ref{t4: mvrv feature} and Fig.~\ref{fig:compare MVRV}), a similar trend is seen, with performance improving across most models when \textit{Asset} and \textit{Rank} features are added. 

In conclusion, incorporating both \textit{Asset} and \textit{Rank} consistently enhances prediction accuracy across all models, as reflected in both the error metrics and the visual alignment with true values.

\section{Related Work}
\label{sec:6_Related_Work}
\subsection{Bitcoin Transaction Network Datasets}
Several datasets have been introduced to analyze the Bitcoin transaction network from various perspectives. These datasets have been valuable in specific contexts.

\textbf{Bitcoin OTC Trust Weighted Signed Network} focuses on trust relationships within a Bitcoin over-the-counter (OTC) marketplace. It is a social network dataset where nodes represent users, and edges represent trust ratings between them. However, this dataset does not capture the full transaction network of Bitcoin or its decentralization aspects~\cite{38_}.

\textbf{Elliptic Dataset} is provided by Elliptic, this dataset contains Bitcoin transaction data labeled as illicit or licit. It is mainly used for fraud detection and anti-money laundering research. However, it is limited in scope and does not cover the entire Bitcoin transaction history, and the semantics of address features are not provided~\cite{39_}.

\textbf{Specific Event Dataset}
Many datasets typically only cover a specific time period or a snapshot of the network. Most of them are proposed for a specific event analysis. They provide information about addresses and transactions but lacks a longitudinal analysis of network properties or a focus on decentralization~\cite{40_, 41_, 42_, 43_}.

These datasets have been instrumental in advancing specific areas of research within the Bitcoin ecosystem. However, they often lack the breadth, depth, and focus necessary for a thorough analysis of Bitcoin's decentralization from an asset perspective. Our work addresses this gap by introducing the first comprehensive dataset covering the entire Bitcoin transaction history from the genesis block to the present. Through our systematic investigation, we provide a deeper understanding of Bitcoin's transaction dynamics and decentralization, offering a valuable resource for future research in this domain.

\subsection{BTC decentralization analysis}
There have been research work proposed to investigate decentralization of BTC, some of them focus on network congestion, delay, and other technical details to evaluate the performance of distributed network~\cite{1_, 2_, 28_, 18_, 19_, 21_, 29_}. Others expose the less-than-optimized decentralization in terms of services decision-making in the BTC system~\cite{3_, 27_, 17_, 20_, 22_, 32_}. However, no quantitative evaluation is provided. 
For quantitative analysis, \cite{5_} proposes a centralization factor based on the ratio of uniformity of mining data. 
Their analysis is constrained to the mining reward and does not concern BTC transactions.
\cite{6_} focuses on evaluating the critical value of the number of nodes needed to control over 51\% of the network by using a Nakamoto coefficient.
This work analyzes the decentralization of the BTC network from the perspective of security in the consensus protocol.
\cite{7_, 8_, 9_} describe decentralization through the randomness degree of data. 
They use statistics and information theory, 
and calculate the variance coefficient and information entropy on blocks mined and address balance to quantify the decentralization of the BTC system.
Most above-mentioned work mainly focus on mining data and offers little insight for the transaction network.

Another line of work investigate the relation between BTC with financial activities. 
\cite{26_, 30_} present the relation between some financial malicious activities and blockchain technology and point out risks and regulatory issues as
Bitcoin interacts with conventional financial systems.
\cite{4_} analyzes the occurrence of large BTC transactions at certain points in time. 
\cite{10_, 11_, 31_} conduct detailed graph measurements on the transactions graph, without relating these results with decentralization though.
\cite{13_} has applied a heuristic for de-anonymization and extracted the user’s graph by merging addresses that belong to the same individuals. 
%

\section{Conclusion and Future Work}
\label{sec:7_Conclusion}

In this paper, we presented the first full-history Bitcoin transaction dataset, spanning over 15 years, to analyze the Bitcoin decentralization phase evolution. By integrating network properties with decentralization metrics to improve financial prediction tasks, we underscore the dataset's value for BTC transaction research. Moving forward, we aim to apply these metrics to other blockchain networks and expand our focus to include hash power centralization, offering a more comprehensive view of Bitcoin's decentralization by combining financial and consensus-layer analyses.

\section{Acknowledgments}

This research is supported by the Ministry of Education, Singapore (reference number: 001526-00001, 001508-00001), WEB 3 SECURITY (reference number: 001529-00001),  Industry Alignment Fund – Pre-positioning (IAF-PP), reference number: 001177-00001. Any opinions, findings and conclusions or recommendations expressed in this material are those of the author(s) and do not reflect the views of Ministry of Education, Singapore.

\bibliography{reference}
\bibliographystyle{IEEEtran}
\end{document}